# Imaging in Radiotherapy

*K. Parodi*
Ludwig-Maximilians-Universität München, Munich, Germany

**Abstract**
With the continued evolution of modern radiation therapy towards high-precision delivery of high therapeutic doses to a tumour while optimally sparing surrounding healthy tissue, imaging is becoming a crucial component for identifying the intended target, positioning it properly at the treatment site, and, in more advanced research applications, visualizing the treatment delivery. This contribution reviews the main roles of imaging in modern external beam radiation therapy, with special emphasis on emerging ion beam therapy techniques aimed at exploiting the favourable properties of the interaction of ions with matter to achieve unprecedented ballistic accuracy in dose delivery.

## 1 Introduction

Over the last two decades, modern radiation therapy with external photon beams has evolved considerably, with the introduction of new delivery techniques such as the use of intensity modulation [1] and rotational therapy [2] for spatio-temporal variation of the radiation dose. In addition to photons, other types of ionizing radiation have been explored with the goal of increased dose conformity for better cure and/or reduced toxicity. In particular, ion beam therapy, especially with proton beams, is rapidly emerging as a promising radiation therapy technique owing to its superior ability to concentrate the beam energy in the tumour while better sparing normal tissue and critical organs compared with photons (Fig. 1). In this context, state-of-the-art technologies that can exploit the charged nature of ions by magnetically steering narrow pencil beams of selected energy over the tumour are being introduced into clinical practice to achieve even better conformity of dose delivery [4].

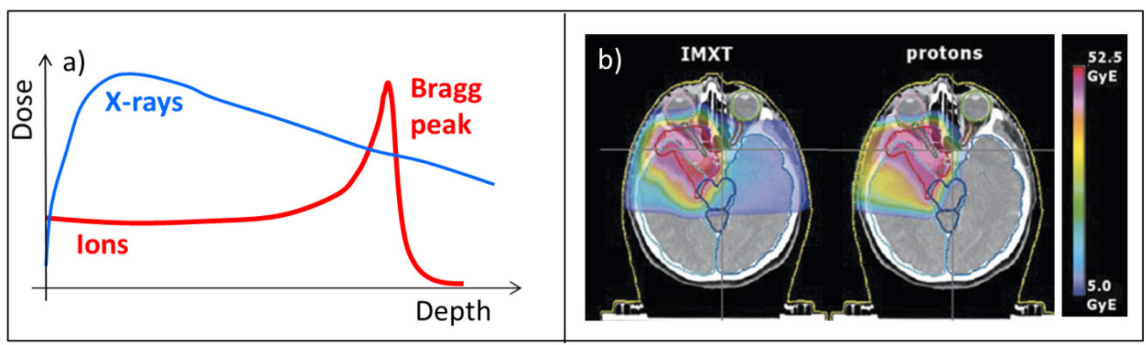

**Fig. 1:** (a) Illustrative representation of depth–dose profiles of photons and ions in water. (b) Treatment plans for a skull-base tumour (adapted from Ref. [3]), comparing the use of photons, delivered with advanced intensity modulation radiation therapy (IMXT), and state-of-the-art treatment with scanned protons, to illustrate the increased tumour–dose conformity of ion therapy due to the characteristic Bragg peak shown in part (a).

The growing ability of modern radiation therapy techniques (using both photon and ion beams) to sculpt the radiation dose tightly to arbitrarily complex tumour shapes (Fig. 1(b)) has tightened the demands on imaging technologies, which play a fundamental role in all three of the stages of treatment planning, treatment delivery, and, in more recent research work, *in vivo* treatment verification aimed at potential plan adaptation, as illustrated in Fig. 2. In the following, the main features of the evolution of imaging technologies for photon and ion radiation therapy will be reviewed, as well as ongoing research, with emphasis on *in vivo* verification of ion beam therapy.

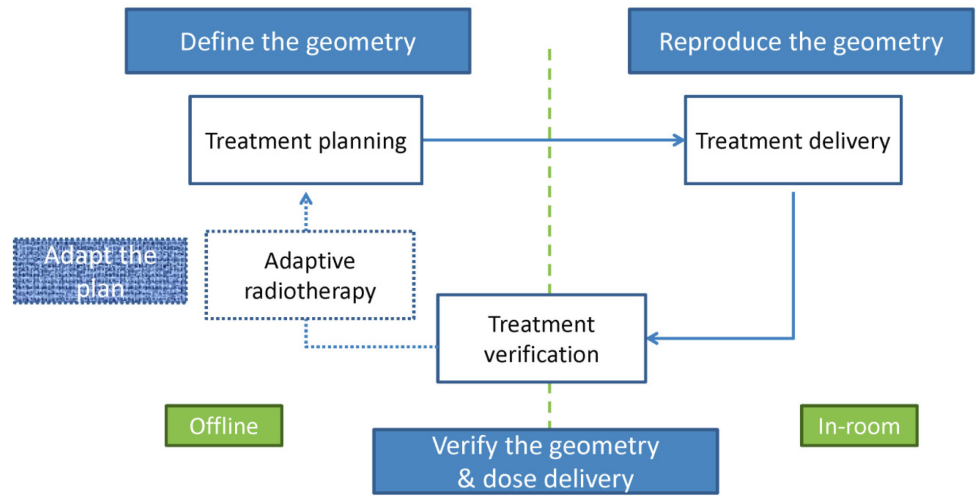

**Fig. 2:** Schematic representation of the role of imaging in the radiation therapy chain, from treatment planning to delivery and verification, with the potential for adaptation. Workflows that must be performed in the treatment room (in-room) are separated by the dashed line from those which can be performed either inside or outside the treatment room (offline) (adapted by courtesy of C. Gianoli, LMU Munich).

## 2 Imaging in radiation therapy

### 2.1 Imaging for treatment planning

The recent physics- and engineering-driven advances in the conformity of dose delivery in radiation therapy were largely motivated by an increased ability to visualize the internal anatomy of the patient with millimetre or submillimetre resolution, as was first made possible by the introduction of volumetric X-ray-based Computed Tomography (CT) in the 1970s [5] and, a few years later, Magnetic Resonance Imaging (MRI) [6]. While X-ray CT still provides the basic patient-specific information for calculating the interaction of radiation with matter for treatment planning, its limited soft-tissue contrast often prevents correct identification of the macroscopic shape of the tumour, called the Gross Target Volume (GTV), for the purpose of delineation. Hence, complementary morphological information from CT and MRI imaging, including physiological motion captured using dedicated time-resolved (4D) acquisition strategies such as respiratory correlated 4D-CT for specific anatomical regions, is typically used to identify the Internal Target Volume (ITV) and Planning Target Volume (PTV). The latter is defined in such a way as to encompass the microscopic extent of the tumour (called the Clinical Target Volume, CTV) expanded by appropriate safety margins, to guarantee coverage of the tumour in the presence of various sources of uncertainty in the treatment.

Beyond morphology, tumour identification increasingly relies also on metabolic and biological features, as assessed by functional imaging via visualization of the spatio-temporal accumulation of injected tracers. In nuclear medicine, Single-Photon Emission Computed Tomography (SPECT) and, more frequently, Positron Emission Tomography (PET) are employed to detect the decay of radionuclides that label selected molecules, resulting in the emission of energetic (a few hundred keV) single photons or annihilation gamma rays [7]. Additional functional information can be provided by special sequences of MRI images that are capable of highlighting physiological processes in the microenvironment of a tumour, as has recently been investigated [8]. Besides contributing to a reduction in inter-observer variability in the delineation of the relevant target volume for treatment planning (Fig. 3) [9], functional imaging opens up the possibility of considering an additional dimension via the definition of the Biological Target Volume (BTV), which describes the complex, heterogeneous microenvironment of the tumour, including subareas of different radiosensitivity [10].

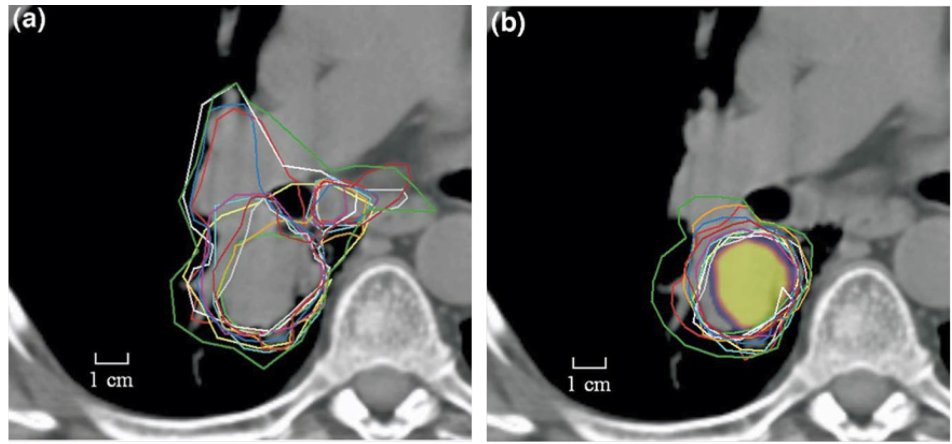

**Fig. 3:** PET data from an $^{18}$F-labelled fluorodeoxyglucose (FDG) tracer, combined with CT imaging, can reduce inter-observer variability in tumour delineation. The coloured contours were defined independently by different radiation oncologists using CT imaging alone in (a) and using PET data in (b), where improved agreement can be seen (reproduced from Ref. [9] with permission.)

Hence, modern radiation therapy is making increasing use of images obtained from different modalities, co-registered using software or hardware fusion (in the case of state-of-the-art combined imaging devices such as PET/CT and PET/MRI scanners), to provide complementary information to the physician to help in the challenging task of segmentation of the tumour and organs at risk, as well as in identification of radioresistant subareas of a tumour which might require a locally enhanced radiation dose for more successful tumour eradication. A treatment plan is then generated using computational engines that optimize the intended dose, in terms of both prescription of the dose to the tumour and constraints on the dose to critical organs, according to the structures identified in the multimodal images used. In this process, the use of CT information is necessary to adapt the well-characterized properties of the interaction of the beam with water to a patient-specific scenario, using the relative electron density (or relative stopping power in the case of ions) extracted from the measured Hounsfield Units (HU) or CT numbers.

## 2.2 Imaging for treatment delivery

Successful delivery of the planned treatment requires correct replication of the patient's position and anatomy to reproduce properly the conditions used in the dose calculation, which is based on images acquired from the patient (as described in Section 2.1) immobilized in the treatment position typically days or weeks before the start of the fractionated therapeutic course of treatment. Although external alignment based on lasers and skin markers or other optical sensors can provide an initial approximate method of positioning that is easy to implement in practice and does not involve exposure to ionizing radiation, the correct positioning of deep-seated tumours requires imaging of the patient's internal anatomy at the treatment site. To this end, individual orthogonal X-ray radiographs have traditionally been used to identify the tumour position indirectly by the alignment of visible anatomical landmarks, such as bony structures or implanted radio-opaque fiducials. However, the restriction to planar images in which contrast from soft tissue is almost absent severely limits the ability to ensure correct positioning of the tumour volume at the treatment site. Hence, the major improvements which have been achieved in the precision of dose delivery have stimulated advances in dedicated instrumentation for *in situ* image guidance, from the use of simultaneous stereoscopic X-ray projections [11] to the widespread use of volumetric kilovoltage X-ray Cone-Beam Computed Tomography (CBCT) [12] and, very recently, MRI (still limited to only a few installations for photon therapy) [13]. In all cases, advanced image processing is used to compare the acquired images with the intended position in the planning CT images, to provide an appropriate position correction to be executed by a robotic positioning system with three (translational) or six (translational and rotational) degrees of freedom. In this way, it can be better ensured that the intended target is accurately positioned on each day of a fractionated treatment, so that the tumour is properly hit while optimally sparing critical structures.

In addition to providing information for position correction of the patient at the treatment site prior to dose delivery, in a technique typically referred to as Image-Guided RadioTherapy (IGRT), the new volumetric representation of the patient can also be used to assess potential anatomical changes, which may make an adaptation of the initial plan necessary, in a technique called Adaptive RadioTherapy (ART). Such adaptation can be done on the basis of a dosimetric evaluation of the initial treatment plan on the modified anatomy of the patient. However, none of the techniques proposed so far for anatomical image guidance in the treatment position at the exact dose delivery site (i.e., without the need to move the immobilized patient to an in-room CT scanner, as is done in some installations) provides accurate CT numbers for dosimetric calculations. This limitation is especially crucial for particle therapy, since the interaction of the ion beam is extremely sensitive to the stopping properties of the tissue, which influence the finite penetration depth of the beam where the maximum energy deposition occurs (the Bragg peak in Fig. 1(a)). Hence, several methods have recently been proposed to recover HU numbers of quality equivalent to the planning CT images for dose calculation, starting from noisy (mostly due to scattering) X-ray-based images produced in CBCT acquisitions (Fig. 4) (see [14–16] and the references therein), or from properly interpreted (e.g., via manual or threshold-based segmentation) MRI images (e.g., [17]). These methods are being extensively investigated. In this way, the new anatomical information can be used to evaluate the dosimetric consequences of anatomical variations prior to treatment delivery, in order to support a decision about whether a new treatment plan needs to be developed to counteract any relevant changes that may have occurred with respect to the initial planning situation. However, updated anatomical imaging cannot provide information about the actual dose application, as will be discussed in the next section.

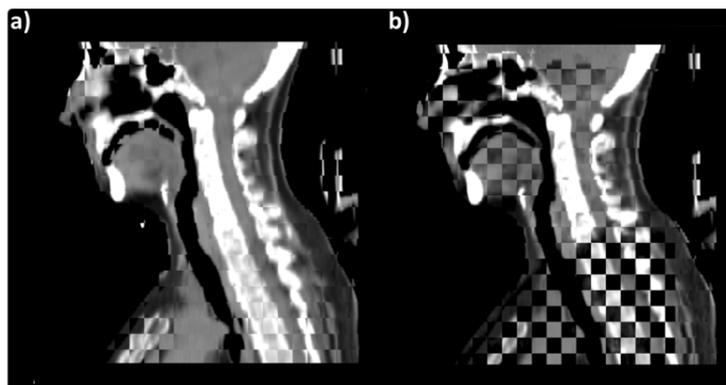

**Fig. 4:** Chequerboard comparison of two intensity-corrected CBCT images, obtained with (a) deformable image registration of the original planning CT image into the CBCT image and (b) a simple HU scaling of the CBCT image based on a population-based look-up table, against a high-quality CT image acquired a few days apart from the CBCT image but weeks after the original planning CT image, in order to capture non-negligible anatomical changes with respect to the planning scenario. Adapted from Ref. [15] with permission.

## 2.3 Imaging for treatment verification

In comparison to photons, ions pose the additional challenge of uncertainty in their range, i.e., inaccuracy in the knowledge of the *in vivo* stopping position of the beam, which determines the position of maximum dose delivery, or Bragg peak. Currently, an intrinsic range uncertainty of 1–3% or even more is associated with the conversion of X-ray CT images to the stopping power ratio of the ions relative to water for calculating the intended treatment [18]; this is in addition to possible set-up errors and anatomical changes (Section 2.2). Although morphological image guidance in clinical ion beam therapy is still predominantly being performed with orthogonal or oblique X-rays, and volumetric (cone-beam or on-rail) CT is just entering routine use in clinical practice, several new methods are being investigated to address the problem of range uncertainty in order to complement or even eventually replace X-ray guidance. Such methods are aimed either at improving the knowledge of the stopping properties of the tissue by using the beam itself for imaging (ion radiography/tomography [19, 20]) or at measuring the stopping position of the ion beam in the target by exploiting secondary emission generated by nuclear reactions (using PET and prompt gamma monitoring [21]), as illustrated in Fig. 5 and described in the following.

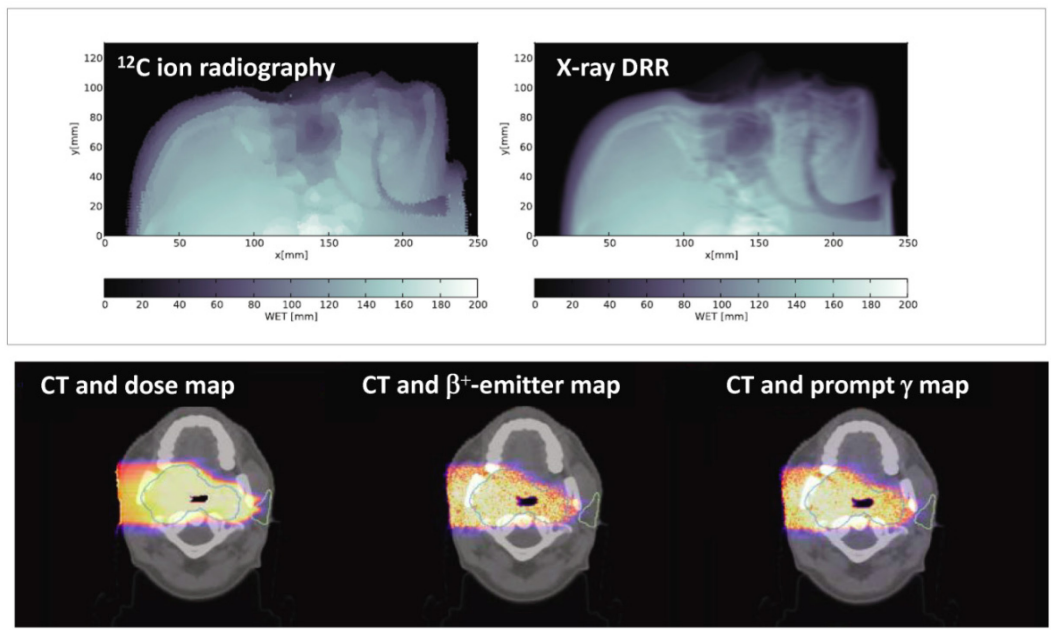

**Fig. 5:** Top: comparison of measured carbon-ion radiography (left) and digitally reconstructed radiography from an X-ray CT image (right) for an anthropomorphic Alderson head phantom, both calibrated to water-equivalent thickness. Adapted from Ref. [22]. (© Institute of Physics and Engineering in Medicine. Reproduced by permission of IOP Publishing. All rights reserved.) Bottom: comparison of CT-based Monte Carlo-calculated positron emitter maps (middle) and prompt gamma distributions (right) for the original plan of a proton treatment of a patient with a head-and-neck tumour (left), all recalculated from an intensity-corrected CBCT image based on the method of Fig. 4(a).

In facilities that are able to produce sufficiently high beam energies for transmission imaging, calibrated ion radiography can provide two-dimensional information about the water-equivalent thickness of the object traversed (Fig. 5, top), which can be used to verify the position of a subject at the treatment site using a dose that is typically lower than that for X-ray images [23], as well as to refine the knowledge of the integral stopping properties of the tissue in order to adjust the planned treatment prior to irradiation [24]. By rotation of the subject or the beam, several projections can be collected to provide tomographic data, yielding volumetric information about the stopping power ratio of the ions in tissue relative to water, which is approximately independent of the beam energy and the ion species. The set-ups proposed for this depend on whether monoenergetic or polyenergetic beams will be used, and include particle trackers for identification of the most likely ion path (especially in the case of the most strongly scattered protons) followed by residual-energy measurements in calorimeters or range telescopes [25], and simpler planar systems for fluence or energy loss measurements of suitably modulated beams [26]. To date, however, no commercial system exists and most research efforts are being devoted to the development of medium-size prototypes, with a focus on clinical application to cranial sites.

In addition to radiographic or tomographic verification of the stopping properties of tissue by ion-based transmission imaging prior to treatment (or even during treatment sessions), nuclear-based methods offer the opportunity to exploit emission signals correlated with the interaction of the beam with tissue during the therapeutic irradiation. To this end, PET imaging has been the method most often investigated clinically, owing to the more mature and readily available detector instrumentation for imaging the transient production of positron emitters in nuclear fragmentation reactions in tissue. Owing to the half-lives of the isotopes formed, ranging from milliseconds or seconds up to 20 minutes, the irradiation-induced PET signal can be measured during or shortly after therapeutic treatment, using instrumentation integrated into the irradiation unit (called in-beam instrumentation) or installed nearby, inside (in-room) or outside (offline) the treatment room [21, 27]. Although the detected PET signal does not have a straightforward correlation with the dose delivered, it can be compared either with predictions based on advanced CT-based computations, using the same HU-range calibration as for treatment planning, or with PET measurements from previous fractions. Clinical experience reported so far with different implementations and ion species indicates the feasibility of validating the HU-range calibration curve *in vivo* [28], as well as of detecting deviations between the planned and actual treatment delivery due to positioning inaccuracies or major anatomical changes [29–31], thus opening up the possibility of adaptation of the treatment before the next treatment delivery (and hence inter-fractional adaptation). However, it has also highlighted methodological challenges, associated mainly with the adaptation of instrumentation originally developed for different applications in diagnostic nuclear medicine or small-animal imaging. Hence, there are ongoing developments aimed especially at new-generation in-beam PET solutions, which rely either on limited-angle dual-head instrumentation exploiting ultrafast time-of-flight (TOF) detectors [32, 33] or on innovative full-ring designs such as the axially shifted 'openPET' [34], together with improvements in computational modelling and data analysis.

Despite the promising clinical results reported by some groups, PET-based verification is intrinsically limited by the delayed emission from the radioactive decay and by the physiological washout that occurs in the time that elapses between production of the positron emitter and detection [21, 27], unless extremely short-lived (millisecond) emitters are used with ultrafast in-beam TOF-PET scanners for quasi-real-time imaging, as has been proposed recently [35, 36]. Conversely, prompt gamma monitoring provides a real-time indirect assessment of the range of the beam from the depth distribution of energetic photons emitted in very fast (on an approximate scale of nanoseconds or less) de-excitation processes, following inelastic scattering and nuclear reactions induced by ion irradiation. Owing to the high photon energies of several MeV and the need for collimated detection for spatial observation of the prompt-gamma–ion-range correlation, several different detection schemes have been proposed, which range from mechanical collimators seen by single or multiple scintillation detectors to more complex Compton cameras, exploiting electronic collimation making use of Compton kinematics (see Refs. [21, 27] and the references therein). Some additional promising developments are aimed at exploiting spectroscopic information about the nucleus-specific, characteristic prompt gamma emission, together with spatial correlation information captured with mechanical collimation [37], or rely on simplified uncollimated detection with very fast scintillators to exploit only the prompt-gamma–proton-range correlation in the time domain [38]. So far, only a prototype system, featuring a single-slit collimator viewed by scintillator detectors, has been realized by a commercial vendor; this is at the stage of clinical evaluation at a few selected proton therapy facilities [39]. The system is limited to a one-dimensional projection of the detected signal and is thus unable to perform volumetric imaging. Nevertheless, in the first clinical applications to passively scattered proton therapy treatment, it has been reported that it is feasible to detect inter-fractional range variations of a few millimetres by comparing prompt gamma profiles measured at different treatment fractions [40]. These findings were corroborated by additional dosimetric recalculations of the treatment plan using new anatomical representations of the patient captured by an on-rail CT system installed in the treatment room [39], thus supporting the promise of the rapidly emerging modality of prompt gamma monitoring for new possibilities of online treatment verification and, potentially, intra-fractional adaptation.

## 3  Conclusion

With the increasing ability of modern photon and ion radiation therapy to provide highly conformal dose delivery, imaging is becoming of paramount importance in all of the main stages of the therapy chain, from treatment planning to treatment delivery and *in vivo* verification, and will be crucial in potential future inter- and intra-fractional adaptation. This contribution has reviewed the main features of the evolution of imaging in radiation therapy, from novel trends in functional imaging for tumour delineation and new treatment concepts, to integrated devices for *in situ* anatomical confirmation and, in the case of the rapidly emerging field of ion beam therapy, unconventional applications of transmission imaging and nuclear-based techniques for *in vivo* ion range verification.

## Acknowledgements

The author would like to thank Drs Gianoli, Dedes, Kurz, and Landry at LMU Munich for providing material used in the graphical illustrations. In this context, support from DFG (under the MAP and carbon ion computed tomography projects) and BMBF (under the SPARTA project) is gratefully acknowledged.